\documentclass[12pt]{article}
\usepackage{amsmath}
\usepackage{bm}
\usepackage[english]{babel}

\begin{document}
\begin{center}
\begin{large}
{\bf Two-particle system in noncommutative space with preserved rotational symmetry}\\
\end{large}
\end{center}

\centerline { Kh. P. Gnatenko \footnote{E-Mail address: khrystyna.gnatenko@gmail.com}, V. M. Tkachuk \footnote{E-Mail address: voltkachuk@gmail.com} }
\medskip
\centerline {\small \it Ivan Franko National University of Lviv, Department for Theoretical Physics,}
\centerline {\small \it 12 Drahomanov St., Lviv, 79005, Ukraine}

\abstract{
We consider a system of two particles in noncommutative space which is rotationally invariant. It is shown that the coordinates of the center-of-mass position and the coordinates of relative motion satisfy noncommutative algebra with corresponding effective tensors of noncommutativity. The hydrogen atom is studied as a two-particle system. We find the corrections to the energy levels of the hydrogen atom up to the second order over the parameter of noncommutativity.


}

\section{Introduction}

The idea of noncommutative structure of space has a long history. This idea was suggested by Heisenberg and later was formalized by Snyder in his paper \cite{Snyder}. In recent years noncommutativity has received much attention. Such an interest is motivated by the development of String Theory and Quantum Gravity (see, for example, \cite{Witten,Doplicher}).

 Canonical version of noncommutative space is realized with the help of the following commutation relations for coordinates and momenta
  \begin{eqnarray}
[X_{i},X_{j}]=i\hbar\theta_{ij},\label{form101}\\{}
[X_{i},P_{j}]=i\hbar\delta_{ij},\\{}
[P_{i},P_{j}]=0,\label{form10001}{}
\end{eqnarray}
where $\theta_{ij}$ is a constant antisymmetric matrix.
The hydrogen atom (see, for example, \cite{Chaichian,Ho,Chaichian1,Chair,Stern,Zaim2,Adorno,Khodja}), the Landau problem (see, for example, \cite{Nair,Bellucci1,Dayi,Li,Dulat}), quantum mechanical system in a central potential \cite{Gamboa}, classical particle in a gravitational potential \cite{Romero,Mirza} and many other problems have been studied in noncommutative space with canonical version of noncommutativity of coordinates.

An important problem is a many-particle problem. Studies of this problem give the possibility to analyze the properties of a wide class of physical systems in noncommutative space.
The classical problem of many particles in noncommutative space-time was examined in \cite{Daszkiewicz}. In this paper the authors studied the set of N interacting harmonic oscillators and the system of N particles moving in the gravitational field.  In \cite{Jabbari} the two-body system of particles interacting through the harmonic oscillator potential was examined on a noncommutative plane. In \cite{Ho} the problems of noncommutative multiparticle quantum mechanics were studied. The authors considered the case when the particles of opposite charges feel opposite noncommutativity. The system of two charged quantum particles was considered in a space with coordinates noncommutativity in \cite{Bellucci}. Also a two-particle system was considered in the context of noncommutative quantum mechanics, characterized by coordinate noncommutativity and momentum noncommutativity in \cite{Djemai}.  The quantum model of many particles moving in twisted N-enlarged Newton-Hooke space-time was proposed in \cite{ Daszkiewicz1}.  As an example the author examined the system of N particles moving in and interacting by the Coulomb potential.

Composite system was studied also in deformed space with minimal length $[X,P]=i\hbar(1+\beta P^{2})$ in \cite{Quesne, Tkachuk}. The authors concluded that the coordinates of the center-of-mass position and the total momentum satisfy noncommutative algebra with an effective parameter of deformation. Using this result the condition for the recovering of the equivalence principle was found  \cite{Tkachuk}.

 It is interesting to note that deformation of algebra of creation and annihilation operators can be used to describe systems of particles possessing interaction and compositeness (see, for example, recent papers \cite{Gavrilik, Gavrilik1} and references therein).

In our previous papers \cite{Gnatenko1, Gnatenko2} we studied composite system in a two-dimensional space with canonical noncommutativity of coordinates $[X_1,X_2]=i\hbar\theta$ where $\theta$ is a constant. We showed that in order to describe motion of a composite system in noncommutative space we have to introduce an effective parameter of noncommutativity. The motion of a composite system was considered in the gravitational field and the equivalence principle was studied. We proposed the condition to recover the equivalence principle in noncommutative space with canonical noncommutativity of coordinates \cite{Gnatenko1}.

In a two-dimensional space with canonical noncommutativity of coordinates the rotational
symmetry is survived.  Note, however, that in a three-dimensional noncommutative space we face the problem of the rotational symmetry breaking \cite{Chaichian,Balachandran1}. To preserve this symmetry new classes of noncommutative algebras were explored (see, for example, \cite{Gnatenko, Kupriyanov} and references therein). Much attention has also been received to the problem of violation of the Lorentz invariance (see, for instance, \cite{Carlson, Morita, Kase}), studying of the the Lorentz symmetry deformations (see, for instance, \cite{Borowiec, Lukerski}, and references therein).

In our paper \cite{Gnatenko} we proposed to construct rotationally invariant noncommutative algebra with the help of generalization of the matrix of noncommutativity to a tensor which is constructed with the help of additional coordinates governed by a rotationally symmetric system. The hydrogen atom energy levels were studied in the rotationally invariant noncommutative space \cite{Gnatenko,Gnatenko3}.

In general case different particles may feel noncommutativity with different tensors of noncommutativity. Therefore there is a problem of describing of motion of a system of particles in noncommutative space.
In present paper we study a system of two particles in rotationally invariant noncommutative space proposed in \cite{Gnatenko}. We find the total momentum of the system and introduce the coordinates of the center-of-mass position. It is shown that the coordinates of the center-of-mass position and the relative coordinates satisfy noncommutative algebra with corresponding effective tensors of noncommutativity. As an example of two-particle system we consider the hydrogen atom. We find the corrections to the energy levels of the atom caused by the noncommutativity of coordinates.

The paper is organized as follows. In Section \ref{rozd2}, a noncommutative space which is rotationally invariant is considered. In Section \ref{rozd3}, we study a two-particle system in rotationally invariant noncommutative space. The total momentum is found and the corresponding coordinates of the center-of-mass position are introduced. In Section \ref{rozd4}, we consider the hydrogen atom as a two-particle system and find the corrections to the energy levels up to the second order in the parameter of noncommutativity. Conclusions are presented in Section \ref{rozd5}.

\section{Rotational symmetry in noncommutative space}\label{rozd2}

 In order to solve the problem of rotational symmetry breaking in noncommutative space in our paper \cite{Gnatenko} we proposed a generalization of the constant antisymmetric matrix $\theta_{ij}$ to a tensor constructed with the help of additional coordinates. We considered the additional coordinates to be governed by a rotationally symmetric system. For simplicity we supposed that the coordinates are governed by the harmonic oscillator \cite{Gnatenko}. Therefore the following rotationally invariant noncommutative algebra was constructed
\begin{eqnarray}
[X_{i},X_{j}]=i\hbar\theta_{ij},\label{form131}\\{}
[X_{i},P_{j}]=i\hbar\delta_{ij},\\{}
[P_{i},P_{j}]=0,\label{form13331}{}
\end{eqnarray}
where tensor of noncommutativity $\theta_{ij}$ is defined as follows
  \begin{eqnarray}
 \theta_{ij}=\frac{\alpha l^2_{P}}{\hbar}\varepsilon_{ijk}\tilde{a}_{k}. \label{form130}
 \end{eqnarray}
Here $\alpha$ is a dimensionless constant, $l_{P}$ is the Planck length, $\tilde{a}_{i}$ are additional dimensionless coordinates governed by the harmonic oscillator
 \begin{eqnarray}
 H_{osc}=\hbar\omega\left(\frac{(\tilde{p}^{a})^{2}}{2}+\frac{\tilde{a}^{2}}{2}\right).\label{form104}
 \end{eqnarray}
The frequency of harmonic oscillator $\omega$ is considered to be very large \cite{Gnatenko}. Therefore, the distance between the energy levels of harmonic oscillator is very large too. So, harmonic oscillator put into the ground state remains in it.

The coordinates $\tilde{a}_{i}$ and the momenta $\tilde{p}^{a}_{i}$ satisfy the following commutation relations
 \begin{eqnarray}
[\tilde{a}_{i},\tilde{a}_{j}]=0,\\{}
[\tilde{a}_{i},\tilde{p}^{a}_{j}]=i\delta_{ij},\\{}
[\tilde{p}^{a}_{i},\tilde{p}^{a}_{j}]=0.{}
 \end{eqnarray}
In addition, coordinates $\tilde{a}_{i}$ commute with $X_{i}$ and $P_{i}$. As a consequence, tensor of noncommutativity (\ref{form130})
also commutes with $X_{i}$ and $P_{i}$. So, $X_{i}$, $P_{i}$ and $\theta_{ij}$ satisfy the same commutation relations as in the case of the canonical version of noncommutativity, moreover algebra (\ref{form131})-(\ref{form13331}) is rotationally invariant.

We can represent the coordinates $X_i$ and the momenta $P_i$ by coordinates $x_i$ and momenta $p_i$ which satisfy the ordinary commutation relations
\begin{eqnarray}
[x_{i},x_{j}]=0,\\{}
[p_{i},p_{j}]=0,\\{}
[x_{i},p_{j}]=i\hbar\delta_{ij}.{}
\end{eqnarray}
Namely, we can use the following representation
\begin{eqnarray}
X_{i}=x_{i}-\frac{1}{2}\theta_{ij}{p}_{j},\label{form01010}\\
P_{i}=p_{i},\label{form01011}
\end{eqnarray}
where $\theta_{ij}$ is given by (\ref{form130}).

The coordinates $x_i$ and the momenta $p_i$ commute with $\tilde{a}_{i}$, $\tilde{p}^a_{i}$, namely  $[x_{i},\tilde{a}_{j}]=0$, $[x_{i},\tilde{p}^a_{j}]=0$, $[p_{i},\tilde{a}_{j}]=0$, $[p_{i},\tilde{p}^a_{j}]=0$.

Explicit representation for coordinates $X_i$ (\ref{form01010}) and momenta $P_i$ (\ref{form01011}) guarantee that the Jacobi identity is satisfied.

  Noncommutative space characterized by the commutation relations (\ref{form131})-(\ref{form13331}), which remain the same after rotation, is rotationally invariant. After rotation $X_{i}^{\prime}=U(\varphi)X_{i}U^{+}(\varphi)$, $\tilde{a}_{i}^{\prime}=U(\varphi)\tilde{a}_{i}U^{+}(\varphi)$, $P_{i}^{\prime}=U(\varphi)P_{i}U^{+}(\varphi)$ we have
\begin{eqnarray}
[X_{i}^{\prime},X_{j}^{\prime}]=i\alpha l^2_{P}\varepsilon_{ijk}\tilde{a}^{\prime}_{k},\\{}
[X_{i}^{\prime},P_{j}^{\prime}]=i\hbar\delta_{ij},\\{}
[P_{i}^{\prime},P_{j}^{\prime}]=0,{}
\end{eqnarray}
here $U(\varphi)=e^{\frac{i}{\hbar}\varphi({\bf n}\cdot{\bf L}^t)}$ is the rotation operator with ${\bf{L}}^t$ being the total angular momentum
\begin{eqnarray}
{\bf L}^t=[{\bf x}\times{\bf p}]+\hbar[{\bf \tilde{a}}\times{\bf \tilde{p}}^{a}].
\end{eqnarray}
 Using (\ref{form130}), (\ref{form01010}) and (\ref{form01011}), the total angular momentum ${\bf{L}^t}$ reads
 \begin{eqnarray}
 {\bf L}^t=[{\bf X}\times{\bf P}]+\frac{\alpha l^2_{P}}{2\hbar}[{\bf P}\times[{\bf \tilde {a}}\times{\bf P}]]+\hbar[{\bf \tilde{a}}\times{\bf\tilde{p}}^{a}].
 \end{eqnarray}
  Therefore algebra (\ref{form131})-(\ref{form13331}) is rotationally invariant.

 Note that ${\bf{L}}^t$ satisfies the following commutation relations
$[X_{i},L^t_{j}]=i\hbar\varepsilon_{ijk}X_{k}$, $[P_{i},L^t_{j}]=i\hbar\varepsilon_{ijk}P_{k}$, $[ \tilde{a}_{i},L^t_{j}]=i\hbar\varepsilon_{ijk} \tilde{a}_{k}$, $[ \tilde{p}^{a}_{i},L^t_{j}]=i\hbar\varepsilon_{ijk} \tilde{p}^{a}_{k}$ which are the same as in the ordinary space.

It is important to mention that from (\ref{form01010}), operators $X_{i}$ depend on the momenta $p_i$ and therefore depend on the mass $m$. It is clear that operators $X_{i}$ do not depend on the mass and can be considered as a kinematic variables in the case when tensor of noncommutativity (\ref{form130}) is proportional to $1/m$, namely when the following condition is satisfied
\begin{eqnarray}
\alpha=\tilde{\gamma}\frac{m_P}{m},\label{form000099}
\end{eqnarray}
where $\tilde{\gamma}$ is a dimensionless constant which is the same for particles of different masses, $m_P$ is the Planck mass.
 It is worth noting that this condition is similar to the condition proposed in \cite{Gnatenko1}. In this paper we considered a two-dimensional space with canonical noncommutativity of coordinates $[X_1,X_2]=i\hbar\theta$ where $\theta$ is a constant and found the following condition
\begin{eqnarray}
\theta=\frac{\gamma}{m}\label{form00009}
\end{eqnarray}
with $\theta$ being the parameter of noncommutativity, which corresponds to the particle of mass $m$, and $\gamma$ is a constant which takes the same value for all particles. The condition gives the possibility to solve an important problem in the two-dimensional noncommutative space, namely the problem of violation of the equivalence principle. Also expression (\ref{form00009}) was derived from the condition of independence of the kinetic energy of a composite system on the composition \cite{Gnatenko1}.

\section{Total momentum and coordinates of the center-of-mass of a two-particle system}\label{rozd3}

Let us consider a system made of two particles of masses $m_{1}$, $m_{2}$ which interact only with each other in rotationally invariant noncommutative space.
 In general case the coordinates of different particles may satisfy noncommutative algebra (\ref{form131})-(\ref{form13331}) with different tensors of noncommutativity. It is also natural to suppose that the coordinates which correspond to different particles commute. Therefore we have the following algebra
 \begin{eqnarray}
[X^{(n)}_{i},X^{(m)}_{j}]=i\hbar\theta^{(n)}_{ij}\delta_{nm},\label{form13111}\\{}
[X^{(n)}_{i},P^{(m)}_{j}]=i\hbar\delta_{ij}\delta_{nm},\\{}
[P^{(n)}_{i},P^{(m)}_{j}]=0.\label{form1333111}{}
\end{eqnarray}
where indices $n$, $m$ label the particles and $\theta^{(n)}_{ij}$ is the tensor of noncommutativity which corresponds to the particle of mass $m_n$.
 We would like to note here that different particles live in the same noncommutative space. Coordinates $a_i$ are responsible for the noncommutativity of a space. Therefore, we suppose that they are the same for different particles. Nevertheless, according to the condition  (\ref{form000099}) particles with different masses feel noncommutativity with different tensors of noncommutativity which depend on their masses. So, taking into account (\ref{form000099}), we can write
\begin{eqnarray}
 \theta^{(n)}_{ij}=\tilde{\gamma}\frac{l_P^2m_P}{\hbar m_n}\varepsilon_{ijk}\tilde{a}_{k}. \label{form13011}
 \end{eqnarray}

 We assume that in noncommutative space Hamiltonian has a similar form as in the ordinary space. Therefore the Hamiltonian which corresponds to the two-particle system is the following
 \begin{eqnarray}
 H_s=\frac{( P^{(1)})^{2}}{2m_{1}}+\frac{( P^{(2)})^{2}}{2m_{2}}+V(|{\bf X}^{(1)}-{\bf X}^{(2)}|),
 \end{eqnarray}
 where $V(|{\bf X}^{(1)}-{\bf X}^{(2)}|)$ is the interaction potential energy of the two particles which depends on the distance between them. Note that coordinates $X^{(n)}_{i}$ satisfy noncommutative algebra (\ref{form13111})-(\ref{form1333111}). We would like to mention that in the ordinary case ($\theta_{ij}=0$) operator of distance corresponds to the ordinary distance. In noncommutative space we have operator of distance $|{\bf X}^{(1)}-{\bf X}^{(2)}|$. Eigenvalues of this operator correspond to the measured distance between two particles in this space. Detailed consideration of this issue is worth to be a subject of separated studies.

In rotationally invariant noncommutative space because of definition of the tensor of noncommutativity (\ref{form130}) we have to take into account additional terms which correspond to the harmonic oscillator. Therefore we consider the total Hamiltonian as follows

\begin{eqnarray}
H=\frac{( P^{(1)})^{2}}{2m_{1}}+\frac{(P^{(2)})^{2}}{2m_{2}}+V(|{\bf X}^{(1)}-{\bf X}^{(2)}|)\nonumber\\+\hbar\omega\left(\frac{(\tilde{p}^{a})^{2}}{2}+\frac{\tilde{a}^{2}}{2}\right).\label{form1113011}
 \end{eqnarray}

Let us introduce the total momentum of  two-particle system as an integral of motion and find the coordinates of the center-of-mass
position as its conjugate variable.
Note that the total momentum ${\bf P}^c$ defined in the traditional way
\begin{eqnarray}
{\bf P}^c={\bf P}^{(1)}+{\bf P}^{(2)}
\label{form4}
\end{eqnarray}
satisfies the following relation
\begin{eqnarray}
[{\bf P}^c,H]=0.{}
\end{eqnarray}
So, the total momentum (\ref{form4}) is an integral of motion in rotationally invariant noncommutative space. The conjugate coordinate to the total momentum reads
\begin{eqnarray}
{\bf X}^c=\frac{m_{1}{\bf X}^{(1)}+m_{2}{\bf X}^{(2)}}{m_{1}+m_{2}}\label{form8888}.
\end{eqnarray}
It is important to mention that the coordinates of the center-of-mass $X^{c}_{i}$ satisfy noncommutative algebra with effective tensor of noncommutativity $\tilde{\theta}_{ij}$. Taking into account (\ref{form13111}), (\ref{form13011}), and (\ref{form8888}) we have
 \begin{eqnarray}
[X^{c}_{i},X^{c}_{j}]=i\hbar\tilde{\theta}_{ij},\label{form0000999}{}
\end{eqnarray}
where effective tensor of noncommutativity is defined as
\begin{eqnarray}
\tilde{\theta}_{ij}=\frac{m_1^2\theta^{(1)}_{ij}+m_2^2\theta^{(2)}_{ij}}{M^2}=\tilde{\gamma}\frac{l_P^2m_P}{\hbar M}\varepsilon_{ijk}\tilde{a}_{k},\label{form500}
\end{eqnarray}
with $M=m_1+m_2$.

The coordinates and the momenta which describe the relative motion can be also introduced in the traditional way
\begin{eqnarray}
{\bf X}^r={\bf X}^{(2)}-{\bf X}^{(1)},\\
{\bf P}^r=\mu_{1}{\bf P}^{(2)}-\mu_{2}{\bf P}^{(1)},\label{form5}
\end{eqnarray}
 with $\mu_{i}=m_{i}/M$.

The coordinates $X^r_i$ and the momenta $P^r_i$ satisfy the following algebra
  \begin{eqnarray}
[X^r_{i},X^r_{j}]=i\hbar(\theta^{(1)}_{ij}+\theta^{(2)}_{ij})=i\hbar\theta^\mu_{ij},\label{form00009990}\\{}
 [X^r_{i}, P^r_{j}]=i\hbar\delta_{ij},\\{}
 [P^r_{i},P^r_{j}]=0.{}
 \end{eqnarray}
So, relative coordinates also satisfy noncommutative algebra with an effective tensor of noncommutativity which is defined as
 \begin{eqnarray}
 \theta^\mu_{ij}=\tilde{\gamma}\frac{l_P^2 m_P}{\hbar\mu}\varepsilon_{ijk}\tilde{a}_k,\label{form5000}
 \end{eqnarray}
 with $\mu=m_{1}m_{2}/(m_{1}+m_{2})$ being the reduced mass. It is worth noting that the coordinates of the center-of-mass position and the coordinates of the relative motion commute due to the condition (\ref{form000099}). We have
\begin{eqnarray}
[X^c_i,X^r_j]=i\frac{\hbar}{M}\left(m_{2}\theta_{ij}^{(2)}-m_{1}\theta_{ij}^{(1)}\right)=0.\label{form9}{}
\end{eqnarray}

The Hamiltonian of two-particle system $H_s$ becomes
\begin{eqnarray}
H_s=\frac{( P^c)^{2}}{2M}+\frac{( P^r)^{2}}{2\mu}+V(|{\bf X}^r|).
\end{eqnarray}

In the next Section as an example of a two-particle system we consider the hydrogen atom.

\section{Effect of coordinate noncommutativity on the energy levels of hydrogen atom}\label{rozd4}

Let us consider the hydrogen atom as two-particle system in rotationally invariant noncommutative space (\ref{form13111})-(\ref{form1333111}). The total Hamiltonian reads
\begin{eqnarray}
 H=\frac{( P^{(1)})^{2}}{2m_{p}}+\frac{(P^{(2)})^{2}}{2m_{e}}-\frac{e^2}{|{\bf X}^{(1)}-{\bf X}^{(2)}|}+\nonumber\\+\hbar\omega\left(\frac{(\tilde{p}^{a})^{2}}{2}+\frac{\tilde{a}^{2}}{2}\right),\label{form777}
\end{eqnarray}
where $m_e$ and $m_p$ are the mass of the electron and the mass of proton, respectively.
According to the results obtained in the previous Section we can rewrite Hamiltonian (\ref{form777}) as follows
\begin{eqnarray}
H=\frac{( P^c)^{2}}{2M}+\frac{(P^r)^{2}}{2\mu}-\frac{e^2}{ X^r}+\hbar\omega\left(\frac{(\tilde{p}^{a})^{2}}{2}+\frac{\tilde{a}^{2}}{2}\right),\nonumber\\\label{form27772}
\end{eqnarray}
here $M=m_e+m_p$, $\mu=m_e m_p/(m_{e}+m_{p})$ and $X^r=|{\bf X}^r|$.

Let us find the corrections to the energy levels of the hydrogen atom up to the second order over the parameter of noncommutativity.

Using (\ref{form01010}) and (\ref{form01011}), it is easy to show that the coordinates $X^r_i$ and the momenta $P^r_i$ can be represented as
\begin{eqnarray}
 X^r_{i}= x^r_{i}+\frac{\tilde{\gamma} m_P l_P^2}{2\hbar\mu}[{\bf\tilde{a}}\times{\bf p}^r]_i,\label{form201010}\\
 P^r_{i}= p^r_{i},\label{form201011}
\end{eqnarray}
where
\begin{eqnarray}
x^r_i=x_i^{(2)}- x_i^{(1)},\\
p^r_i=\mu_{1}p_i^{(2)}-\mu_{2}p_i^{(1)}.\label{form55}
\end{eqnarray}
The coordinates $x^r_i$ and the momenta $p^r_i$ satisfy the ordinary commutation relations
\begin{eqnarray}
[x^r_{i},x^r_{j}]=0,\\{}
[p^r_{i},p^r_{j}]=0,\\{}
[x^r_{i},p^r_{j}]=i\hbar\delta_{ij}.{}
\end{eqnarray}
Using  representation (\ref{form201010}), (\ref{form201011}) and taking into account (\ref{form01011}), we can rewrite Hamiltonian (\ref{form27772}) as follows
\begin{eqnarray}
H=\frac{( p^c)^{2}}{2M}+\frac{( p^r)^{2}}{2\mu}-\frac{e^2}{\left|{\bf x}^r+\frac{\tilde{\gamma} m_P l_P^2}{2\hbar\mu}[\tilde{{\bf a}}\times{\bf p}^r]\right|}+\nonumber\\+ \hbar\omega\left(\frac{(\tilde{p}^{a})^{2}}{2}+\frac{\tilde{a}^{2}}{2}\right),\label{form2777}
\end{eqnarray}
where ${\bf p}^c={\bf p}^{(1)}+{\bf p}^{(2)}$.

Let us write the expansion for $H$ up to the second order in
\begin{eqnarray}
{\bm{\theta}^\mu}=\tilde{\gamma}\frac{l_P^2 m_P}{\hbar\mu}\tilde{{\bf a}}.
\end{eqnarray}
To do that, first let us find the expansion for $X^r$.
 \begin{eqnarray}
   X^r=\left|{\bf x}^r+\frac{\tilde{\gamma} m_P l_P^2}{2\hbar\mu}[\tilde{{\bf a}}\times{\bf p}^r]\right|=\nonumber\\=\sqrt{ (x^r)^{2}-({\bm{\theta}^\mu}\cdot{\bf L}^r)+\frac{1}{4}[{\bm{\theta}^\mu}\times{\bf p}^r]^{2}},\label{form1414}
 \end{eqnarray}
where ${\bf L}^r=[{\bf x}^r\times{\bf p}^r]$. It is worth noting that the operators under the square root do not commute. Therefore, let us write the expansion for $X^r$ in the following form with unknown function $f({\bf x}^r)$
 \begin{eqnarray}
 X^r=x^r-\frac{1}{2 x^r}({\bm{\theta}^\mu}\cdot{ \bf L}^r)-\frac{1}{8 (x^r)^{3}}({\bm{\theta}^\mu}\cdot{ \bf L}^r)^{2}+\nonumber\\+\frac{1}{16}\left(\frac{1}{x^r}[{\bm{\theta}^\mu}\times{\bf p}^r]^{2}+[{\bm{\theta}^\mu}\times{\bf p}^r]^{2}\frac{1}{x^r}+(\theta^\mu)^{2}f({\bf x}^r)\right).\nonumber\\\label{form00002}
 \end{eqnarray}
Function $f({\bf x}^r)$ can be found by squaring left- and right-hand sides of equation (\ref{form00002}). We have
\begin{eqnarray}
\frac{\hbar^{2}}{ (x^r)^{4}}[{\bm{\theta}^\mu}\times{\bf x}^r]^{2}- x^r (\theta^\mu)^{2}f({\bf x}^r)=0.\label{form91}
\end{eqnarray}
So, from (\ref{form91}) we can write
\begin{eqnarray}
(\theta^\mu)^{2}f({\bf x}^r)=\frac{\hbar^{2}}{(x^r)^{5}}[{\bm{\theta}^\mu}\times{\bf x}^r]^{2}.\label{form93}
\end{eqnarray}
On the basis of obtained results (\ref{form00002}), (\ref{form93}) it is easy to find the expansion for $ 1/X^r$ and write expansion of Hamiltonian  (\ref{form2777}) up to the second order in ${\bm{\theta}}^\mu$
\begin{eqnarray}
H=H_{0}+V,\label{form41}\\
H_{0}=H_{h}^{(0)}+H_{osc},\label{form9999}
\end{eqnarray}
here
\begin{eqnarray}
H_{h}^{(0)}=\frac{(p^c)^{2}}{2M}+\frac{( p^r)^{2}}{2\mu}-\frac{e^2}{x^r}
\end{eqnarray}
 is the Hamiltonian of hydrogen atom in the ordinary space and $H_{osc}$ is given by (\ref{form104}). Perturbation caused by the noncommutativity of coordinates $V$ reads
\begin{eqnarray}
V=-\frac{e^{2}}{2(x^r)^{3}}({\bm{\theta}^\mu}\cdot{\bf L}^r)-\frac{3e^{2}}{8(x^r)^{5}}({\bm{\theta}^\mu}\cdot{\bf L}^r)^{2}+\nonumber\\+
\frac{e^{2}}{16}\left(\frac{1}{(x^r)^{2}}[{\bm{\theta}^\mu}\times{\bf p}^r]^{2}\frac{1}{x^r}+\frac{1}{x^r}[{\bm {\theta}^\mu}\times{\bf p}^r]^{2}\frac{1}{(x^r)^{2}}\right.+\nonumber\\+\left.\frac{\hbar^{2}}{(x^r)^{7}}[{\bm{\theta}^\mu}\times{\bf x}^r]^{2}\right).
\end{eqnarray}

Let us find the corrections to the energy levels of the hydrogen atom using perturbation theory.
Note that the total momentum ${\bf p}^c$ is an integral of motion. The momentum ${\bf p}^c$ commutes with $H_{h}^{(0)}$ and $H_{osc}$ and may be replaced by its eigenvalue. So, we can consider the eigenvalue
problem for the following Hamiltonian
\begin{eqnarray}
\tilde{H}_{0}=\frac{(p^r)^{2}}{2\mu}-\frac{e^2}{x^r}+H_{osc}.
\end{eqnarray}
The Hamiltonian which corresponds to the relative motion $(p^r)^{2}/2\mu-e^2/x^r$ commutes with $H_{osc}$. Therefore we can write the eigenvalues and the eigenstates of $\tilde{H}_{0}$
\begin{eqnarray}
E^{(0)}_{n,\{n^{a}\}}=-\frac{e^{2}}{2a^\ast_{B}n^{2}}+\hbar\omega\left(n_{1}^{a}+n_{2}^{a}+n_{3}^{a}+\frac{3}{2}\right),\\
\psi^{(0)}_{n,l,m,\{n^{a}\}}=\psi_{n,l,m}\psi^{a}_{n_{1}^{a},n_{2}^{a},n_{3}^{a}},
\end{eqnarray}
where $\psi_{n,l,m}$ are well known eigenfunctions of the hydrogen atom in the ordinary space,  $\psi^{a}_{n_{1}^{a},n_{2}^{a},n_{3}^{a}}$, are the eigenfunctions of the three-dimensional harmonic oscillator $H_{osc}$, and $a^\ast_{B}=\hbar^2/\mu c^2$ is the Bohr radius including the effect of reduced mass.

In the case when the oscillator is in the ground state,
according to the perturbation theory, we have
\begin{eqnarray}
\Delta E^{(1)}_{n,l}=\langle\psi^{(0)}_{n,l,m,\{0\}}|V|\psi^{(0)}_{n,l,m,\{0\}}\rangle=\nonumber\\=
-\frac{\hbar^{2}e^{2}\langle(\theta^\mu)^{2}\rangle}{(a^\ast_{B})^{5}n^{5}}\left(\frac{1}{6l(l+1)(2l+1)}\right.-\nonumber\\\left.-\frac{6n^{2}-2l(l+1)}{3l(l+1)(2l+1)(2l+3)(2l-1)}\right.+\nonumber\\\left.+\frac{5n^{2}-3l(l+1)+1}{2(l+2)(2l+1)(2l+3)(l-1)(2l-1)}\right.-\nonumber\\\left.-\frac{5}{6}\frac{5n^{2}-3l(l+1)+1}{l(l+1)(l+2)(2l+1)(2l+3)(l-1)(2l-1)}\right),\label{form411}
\end{eqnarray}
where $\langle(\theta^\mu)^{2}\rangle$ is given by
\begin{eqnarray}
\langle(\theta^\mu)^{2}\rangle=\langle\psi^{a}_{0,0,0}|\theta_\mu^{2}|\psi^{a}_{0,0,0}\rangle=\frac{3}{2}\left(\frac{\tilde{\gamma}m_Pl^2_{P}}{\hbar\mu}\right)^2.\label{form887}
\end{eqnarray}
The details of calculation of the corresponding integrals can be found in \cite{Gnatenko}.
In the second order of the perturbation theory we have
\begin{eqnarray}
\Delta
E_{n,l,m,\{0\}}^{(2)}=\nonumber\\=\sum_{n^{\prime},l^{\prime},m^{\prime},\{n^{a}\}}\frac{\left|\left\langle\psi^{(0)}_{n^{\prime},l^{\prime},m^{\prime},\{n^{a}\}}\left|
V\right|\psi^{(0)}_{n,l,m,\{0\}}\right\rangle\right|^{2}}{E^{(0)}_{n}-E^{(0)}_{n^{\prime}}-\hbar\omega(n^{a}_{1}+n^{a}_{2}+n^{a}_{3})},\label{form311}
\end{eqnarray}
where the set of numbers $n^{\prime}$, $l^{\prime}$, $m^{\prime}$,
$\{n^{a}\}$ does not coincide with the set $n$, $l$,
$m$, $\{0\}$. We also use the following notation for the unperturbed energy
of the hydrogen atom $E^{(0)}_{n}=-e^{2}/(2a^\ast_{B}n^{2})$ .
In the case of $\omega\rightarrow\infty$ we have
\begin{eqnarray}
\lim\limits_{\omega\rightarrow\infty}\Delta
E_{n,l,m,\{0\}}^{(2)}=0.\label{form300}
\end{eqnarray}

So, the corrections to the energy levels up to the second order in the parameter of noncommutativity are the following
\begin{eqnarray}
\Delta E_{n,l}=\Delta E^{(1)}_{n,l}.\label{form30000}
\end{eqnarray}
It is important to note that the obtained result (\ref{form30000}) is divergent for $ns$ and $np$ energy levels. It means that expansion of $1/X^r$ over the parameter of noncommutativity can not be applied for calculation of the corrections to $ns$ and $np$ energy levels.
Therefore, let us write the corrections to the $ns$ energy levels in the following way
\begin{eqnarray}
\Delta E_{ns}=\left\langle\psi^{(0)}_{n,0,0,\{0\}}\left|\frac{e^{2}}{x^r}-\frac{e^{2}}{ X^r}\right|\psi^{(0)}_{n,0,0,\{0\}}\right\rangle,
 \label{form721}
 \end{eqnarray}
where $X^r$ is given by (\ref{form1414}). Here we do not use the expansion over the parameter of noncommutativity. Note that $(\bm{\theta}^\mu\cdot{\bf L}^r)$ commutes with $[\bm {\theta}^\mu\times{\bf p}^r]^{2}$ and $(x^r)^{2}$. Also, it is important that $(\bm {\theta}^\mu\cdot{\bf L}^r)\psi^{(0)}_{n,0,0,\{0\}}=0$. Therefore, we can rewrite the corrections (\ref{form721}) in the following form
\begin{eqnarray}
\Delta E_{ns}=\nonumber\\
\left\langle\psi^{(0)}_{n,0,0,\{0\}}\left|\frac{e^{2}}{x^r}-\frac{e^{2}}{\sqrt{(x^r)^{2}+\frac{1}{4}[{\bm{\theta}^\mu}\times{\bf p}^r]^{2}}}\right|\psi^{(0)}_{n,0,0,\{0\}}\right\rangle\nonumber\\=\frac{\chi^{2}e^{2}}{a_{B}}I_{ns}(\chi),\label{form90333}\nonumber\\
 \end{eqnarray}
where we use the notation
 \begin{eqnarray}
I_{ns}(\chi)=\int d\tilde{\bf a} \tilde{\psi}^{a}_{0,0,0}(\tilde{{\bf a}})\int d(\tilde{\bf x}^r)\tilde{\psi}_{n,0,0}(\chi\tilde{\bf x}^r) \left(\frac{1}{\tilde{x}^r}\right.-\nonumber\\ \left.
-\frac{1}{\sqrt{(\tilde{x}^r)^2+[\tilde{{\bf a}}\times\tilde{\bf p}^r]^{2}}}\right)\tilde{\psi}_{n,0,0}(\chi\tilde{\bf x}^r)\tilde{\psi}^{a}_{0,0,0}(\tilde{{\bf a}}),\nonumber\\\label{form947}
 \end{eqnarray}
with
\begin{eqnarray}
 \chi=\sqrt{\frac{\tilde{\gamma}m_P}{2\mu}}\frac{l_P}{a^\ast_B}.\label{form8407}
 \end{eqnarray}
Here $\tilde{\psi}_{n,0,0}(\chi\tilde{\bf x}^r)$ are dimensionless eigenfunctions of the hydrogen atom \begin{eqnarray}
\tilde{\psi}_{n,0,0}(\chi\tilde{\bf x}^r)=\sqrt{\frac{1}{\pi n^{5}}}e^{-\frac{\chi \tilde{ x}^r}{n}}L_{n-1}^{1}\left(\frac{2\chi  \tilde{x}^r}{n}\right)
  \end{eqnarray}
with $L_{n-1}^{1}\left(\frac{2\chi \tilde{x}^r}{n}\right)$ being the generalized Laguerre polynomials,
\begin{eqnarray}
\tilde{\psi}^{a}_{0,0,0}(\tilde{{\bf a}})=\pi^{-\frac{3}{4}}e^{-\frac{\tilde{{a}}^2}{2}}
\end{eqnarray}
 are the dimensionless eigenfunctions corresponding to the harmonic oscillator,
 \begin{eqnarray}
 {\tilde{\bf x}}^r=\sqrt{\frac{2\mu}{\tilde{\gamma}m_P l_P^2}}{\bf x}^r.
\end{eqnarray}
 Integral (\ref{form947}) has a finite value in the case of $\chi=0$. Consequently, for $\chi\rightarrow0$ the asymptotic of $\Delta E_{ns}$ reads
 \begin{eqnarray}
\Delta E_{ns}=\frac{\chi^{2}e^{2}}{a^\ast_{B}}I_{ns}(0)=\frac{\chi^{2}e^{2}}{a^\ast_{B}}\langle I_{ns}(0,\tilde{{\bf a}})\rangle_{\tilde{{\bf a}}},\label{form930}
 \end{eqnarray}
where $\langle...\rangle_{\tilde{{\bf a}}}$ denotes $\langle\tilde{\psi}^{a}_{0,0,0}(\tilde{{\bf a}})|...|\tilde{\psi}^{a}_{0,0,0}(\tilde{{\bf a}})\rangle$. Let us consider the integral
 \begin{eqnarray}
I_{ns}(\chi,\tilde{{\bf a}})=\int d(\tilde{\bf x}^r)\tilde{\psi}_{n,0,0}(\chi \tilde{\bf x}^r) \left(\frac{1}{\tilde{x}^r}\right.-\nonumber\\ \left.
-\frac{1}{\sqrt{(\tilde{x}^r)^2+[\tilde{{\bf a}}\times\tilde{\bf p}^r]^{2}}}\right)\tilde{\psi}_{n,0,0}(\chi \tilde{\bf x}^r).\label{form901}
\end{eqnarray}
For $\chi=0$ we have
\begin{eqnarray}
I_{ns}(0,\tilde{{\bf a}})\simeq1.72\frac{\pi\tilde{a}}{4n^3},\label{form19912}
\end{eqnarray}
 with $\tilde{ a}=|\tilde{{\bf a}}|$. The details of calculation of integral (\ref{form901}) at $\chi=0$ are presented in our previous paper \cite{Gnatenko3}.

So, taking into account (\ref{form8407}), (\ref{form930}), (\ref{form19912}), we have the following result for the leading term in the asymptotic expansion of the corrections to the $ns$ energy levels over the small parameter of noncommutativity
\begin{eqnarray}
\Delta E_{ns}\simeq1.72\frac{\hbar\langle\theta^\mu\rangle\pi e^{2}}{8(a^\ast_{B})^{3}n^{3}},\label{form7999}
\end{eqnarray}
where
\begin{eqnarray}
\langle\theta^\mu\rangle=\langle\psi^{a}_{0,0,0}|\theta^\mu|\psi^{a}_{0,0,0}\rangle=\frac{ 2\tilde{\gamma}m_Pl^2_{P}}{\sqrt{\pi}\hbar\mu}.
\end{eqnarray}

It is important to note that $ns$ energy levels are more sensitive to the noncommutativity of coordinates.  Corrections to these levels  are proportional to $\langle\theta^\mu\rangle$ (\ref{form7999}). For the corrections to the energy levels with $l>1$ we obtained that they are proportional to $\langle(\theta^\mu)^{2}\rangle$ (\ref{form30000}). Corrections to the $np$ energy levels are  planned to be considered in a forthcoming publication.

\section{Conclusions}\label{rozd5}

In this paper we have studied a two-particle problem in noncommutative space with preserved rotational symmetry proposed in \cite{Gnatenko}. We have considered rotationally invariant noncommutative algebra which is constructed with the help of generalization of the matrix of noncommutativity to a tensor constructed with the help of additional coordinates (\ref{form130}).
We have studied a general case when different particles satisfy noncommutative algebra with different tensors of noncommutativity (\ref{form13111})-(\ref{form1333111}).  In such a case the total momentum has been found as an integral of motion and the corresponding coordinates of the center-of-mass position have been introduced.  We have concluded that the coordinates of the center-of-mass and the relative coordinates satisfy noncommutative algebra with corresponding effective tensors of noncommutativity (\ref{form0000999}), (\ref{form00009990}). It is important to note that in the case when the masses of the particles of the system are the same, $m_1=m_2$, and $\theta^{(1)}_{ij}=\theta^{(2)}_{ij}=\theta_{ij}$ according to (\ref{form500}), we have that
$\tilde{\theta}_{ij}=\theta_{ij}/2$. So, there is a reduction of the effective tensor of noncommutativity  with respect to the tensor of noncommutativity for the individual particles. Also,  from (\ref{form5000}) we have $\theta^\mu_{ij}=2\theta_{ij}$. So, the relative motion is more sensitive to the noncommutativity of the coordinates comparing to the motion of the individual particles.

The hydrogen atom has been studied as a two-particle system. We have found the corrections to the energy levels of the atom caused by the noncommutativity of coordinates (\ref{form30000}), (\ref{form7999}). On the basis of the obtained results we have concluded that the $ns$ energy levels are more sensitive to the noncommutativity of the coordinates.

\section*{Acknowledgments}
This work was partly supported by Project FF-30F (No. 0116U001539) from the
Ministry of Education and Science of Ukraine and by the grant from the State Fund For Fundamental Research of Ukraine, Competition F-64, Project "Classical and quantum systems beyond standard approaches".

\end{document}